\documentclass[%
 aps,
 superscriptaddress,
 prl,
 amsmath,amssymb,amsfonts,
 reprint,showpacs
]{revtex4-1}

\usepackage{graphicx}
\usepackage{dcolumn}
\usepackage{bm}
\usepackage{SIunits}
\usepackage{color}

\usepackage{verbatim}   
\usepackage{ifthen}
\usepackage{isomath}
\usepackage{upgreek}

\usepackage{lineno}
\usepackage{subfigure}
\usepackage[english]{babel}
\usepackage{braket}

\definecolor{mygrey}{gray}{0.35}
\definecolor{myblue}{rgb}{0.2,0.2,0.8}
\definecolor{myzard}{cmyk}{0,0,0.05,0}
\definecolor{mywhite}{rgb}{1,1,1}
\definecolor{myred}{rgb}{1,0.,0.3}

\usepackage[colorlinks=true,citecolor=myblue,linkcolor=myred]{hyperref}

\newboolean{mdebug} 
\setboolean{mdebug}{false} 

{%
	\ifthenelse{\boolean{mdebug}}%
	{%
		\color{red}%
		\begin{tiny}%
		\begin{itemize}%
		\item #1%
		\begin{itemize}%
	}%
	{%
		\expandafter\comment%
	}%
}%
{%
	\ifthenelse{\boolean{mdebug}}%
	{%
		\end{itemize}%
		\end{itemize}%
		\end{tiny}%
		\color{black}%
	}%
	{%
		\expandafter\endcomment
	}%
}%

\begin{document}

\preprint{AIP/123-QED}

\title[Spin-boson model with an engineered reservoir in circuit quantum electrodynamics]{Spin-boson model with an engineered reservoir in circuit quantum electrodynamics}

\author{M. Haeberlein}
 \email{max.haeberlein@wmi.badw-muenchen.de}
 \affiliation{Walther-Mei{\ss}ner-Institut, Bayerische Akademie der Wissenschaften, D-85748 Garching, Germany}
 \affiliation{Physik-Department, Technische Universit{\"a}t M{\"u}nchen, D-85748 Garching, Germany}

\author{F. Deppe}
 \affiliation{Walther-Mei{\ss}ner-Institut, Bayerische Akademie der Wissenschaften, D-85748 Garching, Germany}
 \affiliation{Physik-Department, Technische Universit{\"a}t M{\"u}nchen, D-85748 Garching, Germany}
 \affiliation{Nanosystems Initiative Munich (NIM), Schellingstra{\ss}e 4, 80799 M{\"u}nchen, Germany}

\author{A. Kurcz}
 \affiliation{Instituto de F{\'i}sica Fundamental, IFF-CSIC, Calle Serrano 113b, Madrid E-28006, Spain}

\author{J. Goetz}
 \affiliation{Walther-Mei{\ss}ner-Institut, Bayerische Akademie der Wissenschaften, D-85748 Garching, Germany}
 \affiliation{Physik-Department, Technische Universit{\"a}t M{\"u}nchen, D-85748 Garching, Germany}

\author{A. Baust}
 \affiliation{Walther-Mei{\ss}ner-Institut, Bayerische Akademie der Wissenschaften, D-85748 Garching, Germany}
 \affiliation{Physik-Department, Technische Universit{\"a}t M{\"u}nchen, D-85748 Garching, Germany}
 \affiliation{Nanosystems Initiative Munich (NIM), Schellingstra{\ss}e 4, 80799 M{\"u}nchen, Germany}

\author{P. Eder}
 \affiliation{Walther-Mei{\ss}ner-Institut, Bayerische Akademie der Wissenschaften, D-85748 Garching, Germany}
 \affiliation{Physik-Department, Technische Universit{\"a}t M{\"u}nchen, D-85748 Garching, Germany}
 \affiliation{Nanosystems Initiative Munich (NIM), Schellingstra{\ss}e 4, 80799 M{\"u}nchen, Germany}

\author{K. Fedorov}
 \affiliation{Walther-Mei{\ss}ner-Institut, Bayerische Akademie der Wissenschaften, D-85748 Garching, Germany}
 \affiliation{Physik-Department, Technische Universit{\"a}t M{\"u}nchen, D-85748 Garching, Germany}

\author{M. Fischer}
 \affiliation{Walther-Mei{\ss}ner-Institut, Bayerische Akademie der Wissenschaften, D-85748 Garching, Germany}
 \affiliation{Physik-Department, Technische Universit{\"a}t M{\"u}nchen, D-85748 Garching, Germany}

\author{E.P. Menzel}
 \affiliation{Walther-Mei{\ss}ner-Institut, Bayerische Akademie der Wissenschaften, D-85748 Garching, Germany}
 \affiliation{Physik-Department, Technische Universit{\"a}t M{\"u}nchen, D-85748 Garching, Germany}

\author{M. J. Schwarz}
 \affiliation{Walther-Mei{\ss}ner-Institut, Bayerische Akademie der Wissenschaften, D-85748 Garching, Germany}
 \affiliation{Physik-Department, Technische Universit{\"a}t M{\"u}nchen, D-85748 Garching, Germany}
 \affiliation{Nanosystems Initiative Munich (NIM), Schellingstra{\ss}e 4, 80799 M{\"u}nchen, Germany}

\author{F. Wulschner}
 \affiliation{Walther-Mei{\ss}ner-Institut, Bayerische Akademie der Wissenschaften, D-85748 Garching, Germany}
 \affiliation{Physik-Department, Technische Universit{\"a}t M{\"u}nchen, D-85748 Garching, Germany}

\author{E. Xie}
 \affiliation{Walther-Mei{\ss}ner-Institut, Bayerische Akademie der Wissenschaften, D-85748 Garching, Germany}
 \affiliation{Physik-Department, Technische Universit{\"a}t M{\"u}nchen, D-85748 Garching, Germany}
 \affiliation{Nanosystems Initiative Munich (NIM), Schellingstra{\ss}e 4, 80799 M{\"u}nchen, Germany}

\author{L. Zhong}
 \affiliation{Walther-Mei{\ss}ner-Institut, Bayerische Akademie der Wissenschaften, D-85748 Garching, Germany}
 \affiliation{Physik-Department, Technische Universit{\"a}t M{\"u}nchen, D-85748 Garching, Germany}
 \affiliation{Nanosystems Initiative Munich (NIM), Schellingstra{\ss}e 4, 80799 M{\"u}nchen, Germany}

\author{E. Solano}
 \affiliation{Department of Physical Chemistry, University of the Basque Country UPV/EHU, Apartado 644, E-48080 Bilbao, Spain}
 \affiliation{IKERBASQUE, Basque Foundation for Science, Maria Diaz de Haro 3, 48013 Bilbao, Spain}
 \affiliation{Physik-Department, Technische Universit{\"a}t M{\"u}nchen, D-85748 Garching, Germany}

\author{A. Marx} 
 \affiliation{Walther-Mei{\ss}ner-Institut, Bayerische Akademie der Wissenschaften, D-85748 Garching, Germany}

\author{J.J. Garc{\'i}a-Ripoll}
 \affiliation{Instituto de F{\'i}sica Fundamental, IFF-CSIC, Calle Serrano 113b, Madrid E-28006, Spain}

\author{R. Gross}
 \email{rudolf.gross@wmi.badw-muenchen.de}
 \affiliation{Walther-Mei{\ss}ner-Institut, Bayerische Akademie der Wissenschaften, D-85748 Garching, Germany}
 \affiliation{Physik-Department, Technische Universit{\"a}t M{\"u}nchen, D-85748 Garching, Germany}
 \affiliation{Nanosystems Initiative Munich (NIM), Schellingstra{\ss}e 4, 80799 M{\"u}nchen, Germany}

\date{\today}

\begin{abstract}
A superconducting qubit coupled to an open transmission line represents an implementation of the spin-boson model with a broadband environment. 
We show that this environment can be engineered by introducing partial reflectors into the transmission line, allowing to shape the spectral function, $J(\omega)$, of the spin-boson model. 
The spectral function can be accessed by measuring the resonance fluorescence of the qubit, which provides information on both the engineered environment and the coupling between qubit and transmission line. 
The spectral function of a transmission line without partial reflectors is found to be Ohmic over a wide frequency range, whereas a peaked spectral density is found for the shaped environment. 
Our work lays the ground for future quantum simulations of other, more involved, impurity models with superconducting circuits.
\end{abstract}


\maketitle


Quantum impurity models consist of a small, finite-dimensional quantum system, interacting with a larger system, a bath or environment. 
If the focus is set on the impurity, such as in the spin-boson model~\cite{Legget_spinboson}, these models constitute canonical examples of open quantum systems, providing insight into the validity of the Markovian approximation and in a variety of phase transitions such as the localization (i.e., suppression of the impurity tunneling) at large coupling strengths. 
However, such models are also extremely useful from a practical point of view, describing actual dynamics and coherence of ongoing experiments with quantum dots~\cite{hur_cblock}, few-level emitters coupled to plasmonic waveguides~\cite{gonzalez_entangletwoqubits}, or superconducting qubits~\cite{shnirman_noise,wilhelm_decoherence}. 
In this context, a very relevant question is how to suppress or delay the qubit decoherence by means of external controls,~\cite{voila_decoupling,voila_suppression}, especially as the coupling strengths increase.
However, the focus may also shift from the impurity to the environment. 
Then, as in the typical case of the Kondo problem~\cite{wilson_rng}, we find that small impurities dramatically affect the transport properties of the bigger system. 
This idea has found renewed interest in the field of quantum technologies, where few-level systems can be used to control photonic excitations, both in the optical~\cite{chang_singlephotonrouter,tiecke_nanophotonic} and in the microwave regime~\cite{astafiev_fluorescent,wilson_router}, or inducing photon nonlinearities~\cite{hoi_crosskerr}, thus opening the door to all-optical integrated deterministic quantum information processors.

Superconducting circuits are ideal systems for the study of quantum impurity models and their dynamical properties. 
First, superconducting qubits are broadband tunable devices which allow both strong and, potentially, ultrastrong spin-boson coupling~\cite{us_coupling,forn_uscoupling,alex_us}. 
Second, this coupling can be tuned in magnitude and direction~\cite{peropadre_uscoupling,gambetta_purcellprotection,bialczak_tunablecoupler}. Finally, state-of-the-art superconducting circuit technology allows the engineering of various kinds of photonic/bosonic environments: linear~\cite{zueco_mwphotonics,wilhelm_metamaterials}, nonlinear or even quantum~\cite{macha_metamaterials,rakhmanov_metamaterials}. 
This flexibility permits a paradigm shift, where instead of dealing with the spin-boson model and other impurity models in complex, numerically difficult regimes, one can simulate these models by implementing them in the lab. In this way, it is possible to extract both static and dynamic properties. 
This route provides access to the investigation of Kondo resonances~\cite{lehur_kondo}, the study of non-equilibrium phenomena and validation of numerical techniques~\cite{peropadre_dynuscoupling}, the testing of new theoretical methods for describing spin-boson dynamics and phase transitions~\cite{andreas_phasetra1}, or even the exploration of quantum field theories~\cite{sabin_relfields,mezzacapo_relfields}.

In this letter, we report on the experimental characterization of the spin-boson model using a superconducting quantum two-level system (qubit) embedded into a broadband engineered environment, which is realized by a microwave transmission line containing partial reflectors.
We report how a single qubit in a resonance fluorescence experiment is a powerful probe of the bosonic bath, providing us with both the qubit frequency $\omega_{\text{q}}$~\cite{peropadre_dynuscoupling} and the spectral function, $J(\omega)|_{\omega\,{=}\,\omega_{\text{q}}}$. 
Moreover, we show that by introducing partial reflectors in the form of impedance mismatches into the transmission line, it is possible to shape the spectral function from the usual Ohmic case $J(\omega)\,{\propto}\,\omega$ to inhomogeneous distributions. 
Specifically, we study a structured environment that consists of two partial reflectors on each side of the qubit, producing a broadband, Lorentzian environment. 
Tuning the qubit transition frequency over a wide frequency regime, we can derive the engineered spectral function, $\tilde{J}(\omega)$ with properties intermediate between a transmission line and a resonator with highly transparent mirrors (``bad cavity''). 
Finally, regarding the partial reflectors as point scatterers in the spirit of quantum metamaterials~\cite{zueco_mwphotonics}, we also extract the spectral function of the bare transmission line.
Our data confirms that it is Ohmic over a broad frequency range and allows us to extract the Kondo parameter. 

The two-level system in our experiment is a flux qubit~\cite{orlando_fluxqubit,mooij_fluxqubit} built from a superconducting aluminum ring which is interrupted by three Al/AlO$_x$/Al Josephson junctions.
The effective Hamiltonian $H_\text{q}\,{=}\,\hbar\,(\omega_\text{q}/2)\,\sigma_\text{z}$, where $\sigma_{z}$ is a Pauli spin operator.
An external flux $\Phi$ allows us to tune the energy splitting of the qubit, $\hbar\omega_\text{q}\,{=}\,\sqrt{\Delta^2\,+(2\,I_\text{p}\,\delta\Phi)^2}$. Here, the minimum energy $\Delta$ is the qubit energy gap, $I_\text{p}$ is the qubit persistent current, \(\delta \Phi\,\equiv\,\Phi-\Phi_\text{0}/2\), and $\Phi_\text{0}$ is the flux quantum.
The engineered environment is realized by a $10\,\milli\metre$ long on-chip superconducting aluminum coplanar waveguide transmission line with a characteristic impedance $Z_\text{0}\,{=}\,50\,\ohm$, which is represented as $H_\text{bath}\,{=}\,\sum_k\,\hbar\,\omega_k^{\phantom{\dagger}}\,a_k^\dagger\,a_k^{\phantom{\dagger}}$ within the spin-boson model. 
Here, $a_k^\dagger$ and $a_k$ are the bosonic creation and annihilation operators and $\omega_k$ the oscillator frequencies of the bosonic degrees of freedom.
The inductive coupling of the qubit to the engineered environment is realized as a shared conductor segment of width $500\,\nano\metre$ and length $20\,\micro\metre$ between the qubit loop and the center conductor of the coplanar transmission line. 
The corresponding interaction Hamiltonian reads as
\begin{equation}
\label{eq_hamiltonian}
H_\text{int}\,{=}\,\hbar\,\sin\theta\,\sum_k\,\left(g_{k}^{\phantom{\ast}} \sigma^-\,a_k^\dagger\,{+}\,g_k^*\,\sigma^+\,a_k^{\phantom{\dagger}}\right),
\end{equation}
where $\sigma^{-/+}$ are the qubit lowering/raising operators. 
The coupling rate $g_k$ of the qubit to the mode $k$ is scaled by the flux-dependent quantity $\sin\theta\,{\equiv}\,\Delta/\omega_\text{q}$ with $\theta$ being the Bloch angle. 
Qubit and transmission line are fabricated on a silicon dioxide covered silicon wafer.
The wafer is glued into a gold-plated copper box and mounted into a dilution refrigerator to a base temperature of approximately $25\,\milli\kelvin$.
The two interconnections between the on-chip transmission line and the off-chip wiring lead to a weak impedance mismatch at each of the two chip boundaries: the partial reflectors. 

\begin{figure}[tb]
\includegraphics{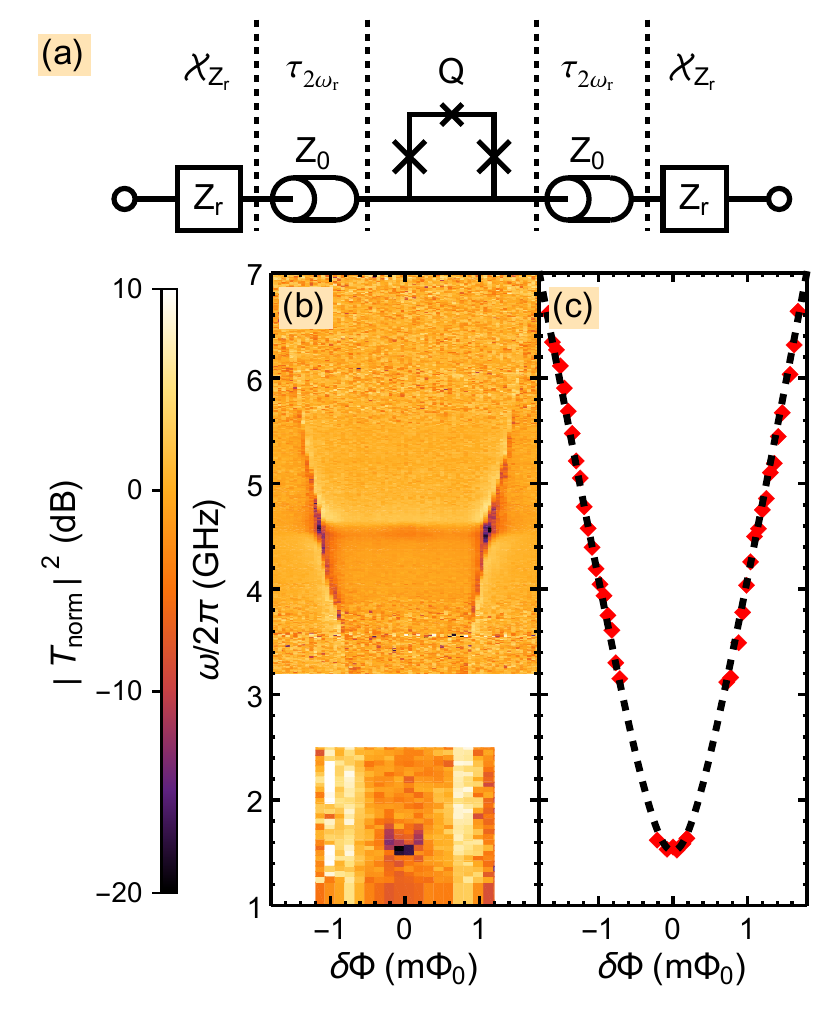}
\caption{(a) The qubit is embedded between two partial reflectors defined by the impedance mismatch $\Delta Z_\text{r} = Z_\text{r} - Z_0$ to the $Z_0\,{=}\,50\,\ohm$ line impedance. In a scattering approach, each segment marked by the dashed lines can be modeled by a transfer matrix. (b) Upper part: Calibrated transmission spectroscopy data. Lower part: Zeeman shift experiment (data scaled by $-30\,\deci\bel$). (c) The red points: Fitted qubit transition frequencies. Dashed curve: Qubit hyperbola with $\Delta\,{=}\,1.52\,\giga\hertz$ and $I_\text{p}\,{=}\,630\,\nano\ampere$.}
\label{fig_hyperbola}
\end{figure}

The total Hamiltonian, $H\,{=}\,H_\text{q}\,{+}\,H_\text{bath}\,{+}\,H_\text{int}$ represents an implementation of the spin-boson model with an engineered reservoir. 
In this model, the interaction between the flux qubit and reservoir is characterized by the spectral function $J(\omega)$. 
In the case of a bare transmission line without reflectors, one expects that the bath contribution to $J(\omega)$ is proportional to $\omega$. 
This situation of a Markovian bath allows one to interpret $J(\omega)|_{\omega=\omega_\text{q}}\,{=}\,\Gamma_1(\omega_\text{q})$ as the spontaneous emission rate of the qubit into a quasi-one dimensional open space. 
In our experiments, we extract $\Gamma_1(\omega)$ by measuring the transmission through the system shown in Fig.~\ref{fig_hyperbola}(a) as a function of the angular frequency $\omega$, and thus can determine the spectral function $J(\omega)$ characterizing the environment. 
To this end, we model the transmittance through our system using transfer matrix theory. 
As detailed in the supplemental material~\cite{supplement_me}, the total $2\times 2$ transfer matrix $\mathcal{T}$ relating the incoming and outgoing modes $a$ and $b$ on one side of the system to the outgoing and incoming modes $a^\prime$ and $b^\prime$ on the other side is given by
\begin{equation}
\label{eq_transfer}
\mathcal{T}(\omega,\omega_\text{q})\,{=}\,\mathcal{X}_{Z_\text{r}} \times \tau_{2\omega_\text{r}}(\omega) \times Q(\omega,\omega_\text{q}) \times \tau_{2\omega_\text{r}}(\omega) \times \mathcal{X}_{Z_\text{r}} \; .
\end{equation}
Here, $\mathcal{X}_{Z_\text{r}}$ is the transfer matrix accounting for the partial reflectors due to the impedance mismatches (they are assumed to be equal), $Q$ describes the flux qubit and $\tau_{2\omega_\text{r}}$ accounts for the propagation in the transmission line between qubit and partial reflectors. 
We assume $\mathcal{X}_{Z_\text{r}}$ to be frequency independent and neglect potential losses in the partial reflectors and the transmission line. 
In this case, the off-diagonal elements of the transfer matrix $\tau_{2\omega_\text{r}}$ vanish and the diagonal ones are $\mathrm{e}^{\pm \imath \varphi}$ with the phase shift $\varphi\,{=}\,\pi\omega/2\omega_{r}$ as discussed in the supplemental material~\cite{supplement_me}. 
Here, $\omega_\text{r}$ is the resonance frequency of the ``bad cavity'' formed in between the partial reflectors. 
With the complex transmission and reflection coefficients $t_{Z_\text{r}}$ and $r_{Z_\text{r}}$ of the partial reflectors~\cite{supplement_me}, the transfer matrices can be written as
\begin{eqnarray*}
Q & = & \left(
    \begin{array}{cc}
    t_\text{q}-r_\text{q}^2/t_\text{q}  & r_\text{q}/t_\text{q} \\
    - r_\text{q}/t_\text{q}      &  1/t_\text{q}
    \end{array}
    \right)
    \\ 
\mathcal{X}_{Z_\text{r}} & = & \left(
    \begin{array}{cc}
    1/t_{Z_\text{r}}^*  & -r_{Z_\text{r}}^*/t_{Z_\text{r}}^* \\
    - r_{Z_\text{r}}/t_{Z_\text{r}}      &  1/t_{Z_\text{r}}
    \end{array}
    \right) \; .
\end{eqnarray*}
Following Ref.~\onlinecite{astafiev_fluorescent} and Ref.~\onlinecite{supplement_me}, $t_\text{q}$ and $r_\text{q}$ are functions of the qubit transition frequency, the qubit dephasing rate $\Gamma_\varphi$, and the qubits spontaneous emission rate $\Gamma_1$ of the qubit into the open transmission line environment. 
As $\mathcal{T}$ is of the same form as $Q$, we obtain a system of equations for the total transmission amplitude $T_\text{t}$.
This amplitude normalized to the case where the qubit is far detuned reads
\begin{equation}
\label{eq_caltra}
T_{\text{norm}}\,{\equiv}\,\frac{T_\text{t}}{T_\text{t}|_{\omega_\text{q}\,{\rightarrow}\,\infty}}\,{=}\,\frac{2\,(\Gamma_\varphi\,{-}\,i\,(\omega\,{-}\,\omega_\text{q}) )}   {\tilde{\Gamma}_\text{1}\,{+}\,2\,\Gamma_\varphi\,{-}\,2\,i\,(\omega\,{-}\,\omega_\text{q})}~,
\end{equation} 
where $\tilde{\Gamma}_\text{1}{\equiv}\,\Gamma_\text{1}\,(1\,{-}\,r_{Z_\text{r}}\,e^{i\,\pi\,\omega/\omega_\text{r}})/(1\,{+}\,r_{Z_\text{r}}\,e^{i\,\pi\,\omega/\omega_\text{r}})$. 
We note that the form of Eq.\,(\ref{eq_caltra}) suggests that the qubit probes a modified environment characterized by a spectral function $\tilde{J}(\omega)\,{=}\,\Re(\tilde{\Gamma_\text{1}}(\omega))$.
In this way, Eq.~\,(\ref{eq_caltra}) allows us to access both the spectral functions $J(\omega)$ and $\tilde{J}(\omega)$ from a single transmission measurement as long as the mismatches are weak enough.

\begin{figure}[tb]
\includegraphics{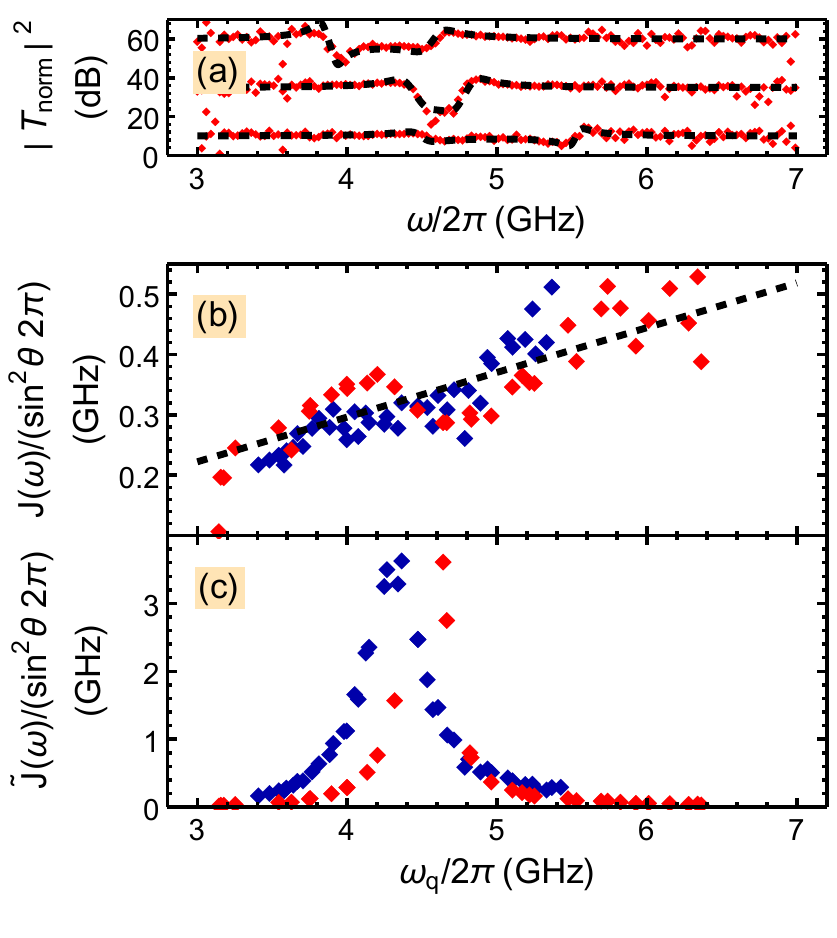}
\caption{%
(a) Calibrated transmission spectroscopy data for $\omega_\text{q}/2\pi = 3.9\,\giga\hertz$ (top), $4.56\,\giga\hertz$ (middle) and $5.5\,\giga\hertz$ (bottom). The data points are displaced by $60\,\deci\bel$, $35\,\deci\bel$, and $10\,\deci\bel$, respectively. Dashed lines: Fits using Eq.~(\ref{eq_caltra}). 
(b) Spectral function $J(\omega)$ for two measurements (blue: $Z_\text{r}\,{=}\,111\,\ohm$; red: $Z_\text{r}\,{=}\,253\,\ohm$) for the bare qubit-transmission-line coupling. Dashed line: Linear fit.
(c) Spectral function $\tilde{J}(\omega)$ for the qubit coupled to the ``bad cavity''-environment [color code as in (b)].
}
\label{fig_coupling}
\end{figure}

Before analyzing the coupling of the qubit to the environment in more detail, we first have to determine the parameters describing the flux qubit.
In the upper part of Fig.\,\ref{fig_hyperbola}(b), we show transmission spectroscopy data at low input power. 
Exemplary fits are shown in Fig.\,\ref{fig_coupling}(a).
In order to identify the qubit energy splitting $\Delta$, which lies outside our amplifier bandwidth, we use the broad resonance as a readout device. 
The results of this standard Zeeman shift experiment~\cite{alex_us} are plotted in the lower part of Fig.~\ref{fig_hyperbola}(b). Fitting Eq.~(\ref{eq_caltra}) to all spectroscopic data, we extract the parameters $\Delta/h\,{=}\,1.52\,\giga\hertz$ and $I_p\,{=}\,630\,\nano\ampere$ determining the flux-dependent qubit transition frequency $\omega(\delta\Phi)$ shown in Fig.~\ref{fig_hyperbola}(c).


In addition to the qubit parameters, Eq.\,(\ref{eq_caltra}) also provides information on the spectral function $J(\omega)$ of the transmission line.
As shown in Fig.\,\ref{fig_coupling}(b), our data directly confirms the expectation that the transmission line presents an Ohmic bath, $J(\omega)/\sin^2\theta\,{=}\,2\,\pi\,\alpha\,\omega$, where the factor $1/\sin^2\theta$ corrects for the qubit contribution to the frequency dependence of the spectral function. 
We extract a Kondo parameter $\alpha\,{\simeq}\,1\,\%$ in the full frequency range of $3\,{-}\,7\,\giga\hertz$ accessible by our experiments.
This result proves the broadband nature of our experiment, clearly distinguishing it from standard quasi-single-mode qubit-resonator physics.
The small value of $\alpha$ justifies our treatment of the transmission in the scattering approach in Eq.\,(\ref{eq_transfer}) and is associated with a mutual inductance of $M\,{\simeq}\,29\,\pico\henry$ between qubit and transmission line~\cite{supplement_me}. 
A mutual inductance of this magnitude corresponds to a single-mode coupling strength $g_k/\omega_k$ of approximately $15\,\%$. 
Such values have already enabled the experiments observation of ultrastrong coupling effects in systems, where qubits were coupled to high-quality resonators. 
In an open line, however, the quantum phase transition to the Kondo regime ($\alpha\,{\ge}\,1/2$) would require to increase the mutual inductance further to above $200\,\pico\henry$~\cite{supplement_me}.
Next, we analyze $J(\omega)$ of our sample for two different, but in each case symmetric partial reflector configurations.
These configurations correspond to the impedance mismatches $\Delta\/Z_\text{r}\,{=}\,Z_\text{r}\,{-}\,Z_\text{0}$, where $Z_\text{r}\,{=}\,\{Z,Z^\prime\}\,{=}\,\{253\,\ohm,\,111\,\ohm\}$ and the associated broad resonances have center frequencies $\omega_\text{r}^{}\,{=}\,\{\omega_0^{},\omega_0^\prime\}\,{=}\,\{4.53\,\giga\hertz,\,4.34\,\giga\hertz\}$.
Requiring that the Kondo parameter of the line should not change for one and the same sample, we obtain the qubit gap $\Delta^\prime/h\,{=}\,1.2\,\giga\hertz$ and the persistent current $I_\text{p}^\prime\,{=}\,644\,\nano\ampere$. 

In the following, we discuss how an engineered reservoir, as we apply it in our experiments, can be used to alter the frequency dependent coupling of the qubit to the transmission line. 
$\tilde{J}(\omega)$ is comprised of a resonator-type contribution, a Lorentzian function characterized by the loss rate $\kappa$, and an Ohmic line contribution $J(\omega)\,{=}\,2\pi\alpha\omega$ characterized by the Kondo parameter $\alpha$.
Our model involves both of these contributions. 
Consequentially, the radiative correction factor applied to the spectral function of the Ohmic line tends towards the Purcell factor for $\kappa\,{\to}\,0$~\cite{supplement_me}.
In contrast to this extreme limit, however, we conduct our experiments in a regime where both contributions matter. 
In this way, we are able to extract the spectral function of the underlying Ohmic line $J(\omega)$ in a broad frequency range.
In Fig.\,\ref{fig_coupling}(c), we show that we can, by varying the reflectance of the partial reflectors, effectively shape the total spectral function $\tilde{J}(\omega)$. 
On the center frequency of the Lorentzian function, $\tilde{J}(\omega_0)/2\,\pi\,\sin^2\theta\,{\simeq}\,8.25\,\giga\hertz$ and  $\tilde{J}(\omega_0^{\prime})/2\,\pi\,\sin^2\theta\,{\simeq}\,2.33\,\giga\hertz$.
Furthermore, from the fits of Eq.~(\ref{eq_caltra}), we obtain $\kappa_\text{tra}/2\pi\,{\simeq}\,185\,\mega\hertz$ and $\kappa_\text{tra}^\prime/2\pi\,{\simeq}\,756\,\mega\hertz$ as the decay rates for the resonances connected to the center frequencies $\omega_0^{}$ and $\omega_0^\prime$, respectively.

\begin{figure}[tb]
\includegraphics{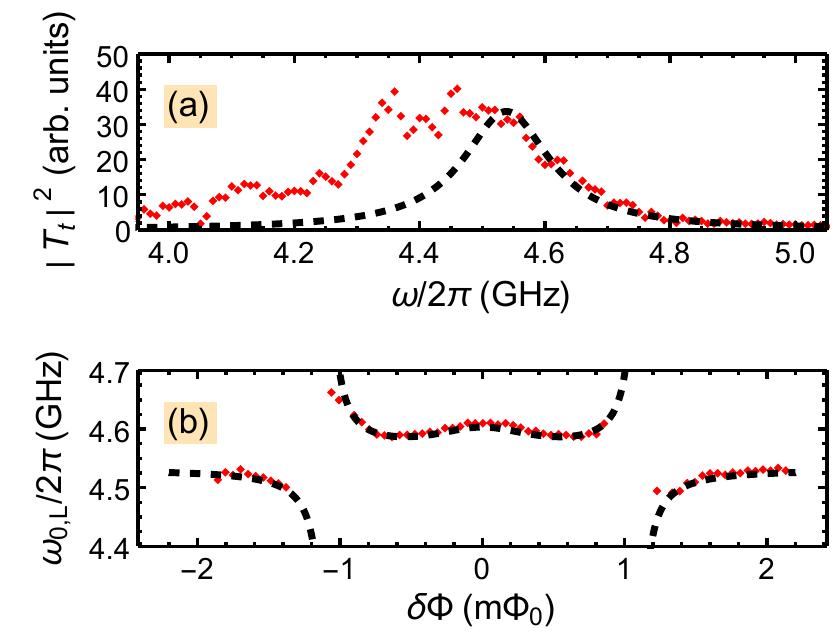}
\caption{(a) Uncalibrated transmission power with the qubit far detuned (diamonds). Dashed line: Lorentzian fit of the resonance. (b) Lamb shift of the ``bad cavity''-resonance. Dashed line: Fit using Eq.~(\ref{eq_resshift}).
}
\label{fig_res}
\end{figure}

Next, since our experiments happen in a regime intermediate between resonator and transmission line physics, we can also analyze our data from a resonator-qubit point of view. 
We are, however, limited by the quality of the Lorentzian fits and therefore only show results for the measurement with $Z_\text{r}\,{=}\,253\,\ohm$.
For this reflector configuration, Fig.\,\ref{fig_res}(a) shows a Lorentzian fit to the uncalibrated transmission spectrum.
The decay rate of the resonator, $\kappa_\text{r}\,{=}\,169\,\mega\hertz$, in the case that the qubit is far detuned agrees well with $\kappa_\text{tra}$ found from the transmission line analysis.
Furthermore, the center frequency of the Lorentzian, $\omega_\text{0,L}$, varies as a function of the flux threading the qubit. As shown in Fig.~\ref{fig_res}(b), this Lamb shift follows the expected behavior~\cite{zueco_lambshift}, 
\begin{equation}
\label{eq_resshift}
\omega_\text{0,L}\,{=}\,\omega_\text{0}\,{+}\,g^2\,(\sin\theta)^2\left(\frac{1}{\omega_\text{q}\,{-}\,\omega_{\text{r}}}\,{+}\,\frac{1}{\omega_\text{q}\,{+}\,\omega_{\text{r}}}\right)~.
\end{equation} 
A numerical fit to the data yields a coupling rate $g\,\sin\theta\,{=}\,216\,\mega\hertz$.
To compare the transmission line coupling with $g$, we calculate (c.f. supplemental material~\cite{supplement_me}) a coupling $g_\text{tra}\,{=}\,233\,\mega\hertz$ the qubit would have to a single mode resonator with resonance frequency $\omega_\text{r}$ made from our transmission line. 
Remarkably, we find a very good agreement between the value of $g$ extracted from a resonator type experiment with the value of $g_\text{tra}$ extracted from the qubit fluorescence.

Finally, we discuss the increase in the spontaneous emission rate of the qubit for our engineered environments. 
In this context, we remark that the expression $\tilde{J}(\omega)\,{=}\,\Re(\tilde{\Gamma_\text{1}}(\omega))$ only holds for Markovian environments where excitations emitted to the environment are not allowed to excite the qubit again. 
Obviously, the latter assertion breaks down with the onset of the strong coupling regime.
Using this, we can by setting $\kappa_{\text{tra}}(\omega_\text{r})\,{=}\,g_\text{tra}(\omega_\text{r})$ define a critical decay rate $\kappa_\text{crit}(\omega_r)$.
With $\kappa_\text{crit}(\omega_0)\,{=}\,233\,\mega\hertz$ and $\kappa_\text{crit}(\omega_0^\prime)\,{=}\,175\,\mega\hertz$, we find that in the first case the Markovian approximation of the bath is violated ($\kappa_\text{tra}\,{\lesssim}\,\kappa_\text{crit}(\omega_0)$). This situation is different in the second case, where $\kappa_\text{tra}^\prime\,{\gg}\,\kappa_\text{crit}(\omega_0^\prime)$, and we can interpret $\tilde{\Gamma_1}/\Gamma_1\,{=}\,\tilde{J}(\omega_0^\prime)/{J}(\omega_0^\prime)\,{\simeq}\,8$ in terms of an increased spontaneous emission.


In summary, we experimentally implement the spin-boson model with an engineered reservoir using superconducting circuits. 
To this end, we study a superconducting flux qubit in a continuous environment modified by two discrete scatterers. 
The formation of a broad resonance effectively allows us to control the density of states in the environment of the qubit. 
In this way, we can increase the spontaneous emission by a factor of $8$. 
Our model of the combined qubit-resonance system additionally supplies information about the spontaneous emission rate of the qubit in a bare transmission line. 
We confirm that the transmission line represents an Ohmic environment and our coupling corresponds to a Kondo parameter $\alpha\,{=}\,1\,\%$. 
Additionally, we crosscheck our model by comparing the resonator Lamb shift for the qubit being far detuned and find a very good agreement to the transmission line coupling.
Our work constitutes an important step towards the quantum simulation of impurity models with superconducting circuits.
In this context, the modeling of the effect of broad resonances on the environment paves the way for protocols where qubits can be coupled to and decoupled from engineered broadband reservoirs dynamically in the spirit of band engineering in photonic crystals.
Finally, we show that increasing the mutual inductance to the technologically challenging but in principle feasible value of $200\,\pico\henry$ is required to reach the Kondo regime with flux qubits.

This work is supported by the German Research
Foundation through SFB 631 and FE 1564/1-1, EU projects CCQED,
PROMISCE and SCALEQIT, the doctorate program ExQM of the Elite Network of Bavaria,
Spanish MINECO Project FIS2012-33022 and CAM Research Network QUITEMAD+.
E. S. acknowledges support from Basque Government IT472-10; Spanish MINECO FIS2012-36673- C03-02; UPV/EHU UFI 11/55; PROMISCE and SCALEQIT EU projects. E.S. acknowledges the hospitality of Walther-Mei{\ss}ner-Institut and Institute for Advanced Study at TUM.

\bibliographystyle{apsrev4-1}
\bibliography{bib}

\begin{thebibliography}{8}
\bibitem [{myf()}]{myfootnote}%
  \BibitemOpen
  \href@noop {} {}\bibinfo {note} {One can show that in the case
  $\Gamma_\varphi \rightarrow 0$ the transfer matrix Q recovers the symmetry of
  Eq.~(\ref{chi}) and hence with $Q \in {\rm SU} (1,1)$ it is $H \in {\rm SU}
  (1,1)$.}\BibitemShut {Stop}%
\end{thebibliography}

\end{document}



\title[Supplemental Material for ``Spin-boson model with an engineered reservoir in circuit quantum electrodynamics'']{Supplemental Material to ``Spin-boson model with an engineered reservoir in circuit quantum electrodynamics''}

\author{M. Haeberlein}
 \email{max.haeberlein@wmi.badw-muenchen.de}
 \affiliation{Walther-Mei{\ss}ner-Institut, Bayerische Akademie der Wissenschaften, D-85748 Garching, Germany}
 \affiliation{Physik-Department, Technische Universit{\"a}t M{\"u}nchen, D-85748 Garching, Germany}

\author{F. Deppe}
 \affiliation{Walther-Mei{\ss}ner-Institut, Bayerische Akademie der Wissenschaften, D-85748 Garching, Germany}
 \affiliation{Physik-Department, Technische Universit{\"a}t M{\"u}nchen, D-85748 Garching, Germany}
 \affiliation{Nanosystems Initiative Munich (NIM), Schellingstra{\ss}e 4, 80799 M{\"u}nchen, Germany}

\author{A. Kurcz}
 \affiliation{Instituto de F{\'i}sica Fundamental, IFF-CSIC, Calle Serrano 113b, Madrid E-28006, Spain}

\author{J. Goetz}
 \affiliation{Walther-Mei{\ss}ner-Institut, Bayerische Akademie der Wissenschaften, D-85748 Garching, Germany}
 \affiliation{Physik-Department, Technische Universit{\"a}t M{\"u}nchen, D-85748 Garching, Germany}

\author{A. Baust}
 \affiliation{Walther-Mei{\ss}ner-Institut, Bayerische Akademie der Wissenschaften, D-85748 Garching, Germany}
 \affiliation{Physik-Department, Technische Universit{\"a}t M{\"u}nchen, D-85748 Garching, Germany}
 \affiliation{Nanosystems Initiative Munich (NIM), Schellingstra{\ss}e 4, 80799 M{\"u}nchen, Germany}

\author{P. Eder}
 \affiliation{Walther-Mei{\ss}ner-Institut, Bayerische Akademie der Wissenschaften, D-85748 Garching, Germany}
 \affiliation{Physik-Department, Technische Universit{\"a}t M{\"u}nchen, D-85748 Garching, Germany}
 \affiliation{Nanosystems Initiative Munich (NIM), Schellingstra{\ss}e 4, 80799 M{\"u}nchen, Germany}

\author{K. Fedorov}
 \affiliation{Walther-Mei{\ss}ner-Institut, Bayerische Akademie der Wissenschaften, D-85748 Garching, Germany}
 \affiliation{Physik-Department, Technische Universit{\"a}t M{\"u}nchen, D-85748 Garching, Germany}

\author{M. Fischer}
 \affiliation{Walther-Mei{\ss}ner-Institut, Bayerische Akademie der Wissenschaften, D-85748 Garching, Germany}
 \affiliation{Physik-Department, Technische Universit{\"a}t M{\"u}nchen, D-85748 Garching, Germany}

\author{E.P. Menzel}
 \affiliation{Walther-Mei{\ss}ner-Institut, Bayerische Akademie der Wissenschaften, D-85748 Garching, Germany}
 \affiliation{Physik-Department, Technische Universit{\"a}t M{\"u}nchen, D-85748 Garching, Germany}

\author{M. J. Schwarz}
 \affiliation{Walther-Mei{\ss}ner-Institut, Bayerische Akademie der Wissenschaften, D-85748 Garching, Germany}
 \affiliation{Physik-Department, Technische Universit{\"a}t M{\"u}nchen, D-85748 Garching, Germany}
 \affiliation{Nanosystems Initiative Munich (NIM), Schellingstra{\ss}e 4, 80799 M{\"u}nchen, Germany}

\author{F. Wulschner}
 \affiliation{Walther-Mei{\ss}ner-Institut, Bayerische Akademie der Wissenschaften, D-85748 Garching, Germany}
 \affiliation{Physik-Department, Technische Universit{\"a}t M{\"u}nchen, D-85748 Garching, Germany}

\author{E. Xie}
 \affiliation{Walther-Mei{\ss}ner-Institut, Bayerische Akademie der Wissenschaften, D-85748 Garching, Germany}
 \affiliation{Physik-Department, Technische Universit{\"a}t M{\"u}nchen, D-85748 Garching, Germany}
 \affiliation{Nanosystems Initiative Munich (NIM), Schellingstra{\ss}e 4, 80799 M{\"u}nchen, Germany}

\author{L. Zhong}
 \affiliation{Walther-Mei{\ss}ner-Institut, Bayerische Akademie der Wissenschaften, D-85748 Garching, Germany}
 \affiliation{Physik-Department, Technische Universit{\"a}t M{\"u}nchen, D-85748 Garching, Germany}
 \affiliation{Nanosystems Initiative Munich (NIM), Schellingstra{\ss}e 4, 80799 M{\"u}nchen, Germany}

\author{E. Solano}
 \affiliation{Department of Physical Chemistry, University of the Basque Country UPV/EHU, Apartado 644, E-48080 Bilbao, Spain}
 \affiliation{IKERBASQUE, Basque Foundation for Science, Maria Diaz de Haro 3, 48013 Bilbao, Spain}
 \affiliation{Physik-Department, Technische Universit{\"a}t M{\"u}nchen, D-85748 Garching, Germany}

\author{A. Marx} 
 \affiliation{Walther-Mei{\ss}ner-Institut, Bayerische Akademie der Wissenschaften, D-85748 Garching, Germany}

\author{J.J. Garc{\'i}a-Ripoll}
 \affiliation{Instituto de F{\'i}sica Fundamental, IFF-CSIC, Calle Serrano 113b, Madrid E-28006, Spain}

\author{R. Gross}
\email{rudolf.gross@wmi.badw-muenchen.de}
 \affiliation{Walther-Mei{\ss}ner-Institut, Bayerische Akademie der Wissenschaften, D-85748 Garching, Germany}
 \affiliation{Physik-Department, Technische Universit{\"a}t M{\"u}nchen, D-85748 Garching, Germany}
 \affiliation{Nanosystems Initiative Munich (NIM), Schellingstra{\ss}e 4, 80799 M{\"u}nchen, Germany}

\date{\today}

\begin{abstract}
The following document provides Supplemental Material regarding the experimental methods,
data analysis, and theoretical background for the manuscript ``Spin boson model with an engineered reservoir in circuit QED''.
\end{abstract}

\pacs{85.25.-j, 85.25.Cp, 85.25.Hv, 81.05.Xj, 78.67.Pt}

\maketitle


\section{Experimental Techniques}

\begin{figure}[h!]
\includegraphics{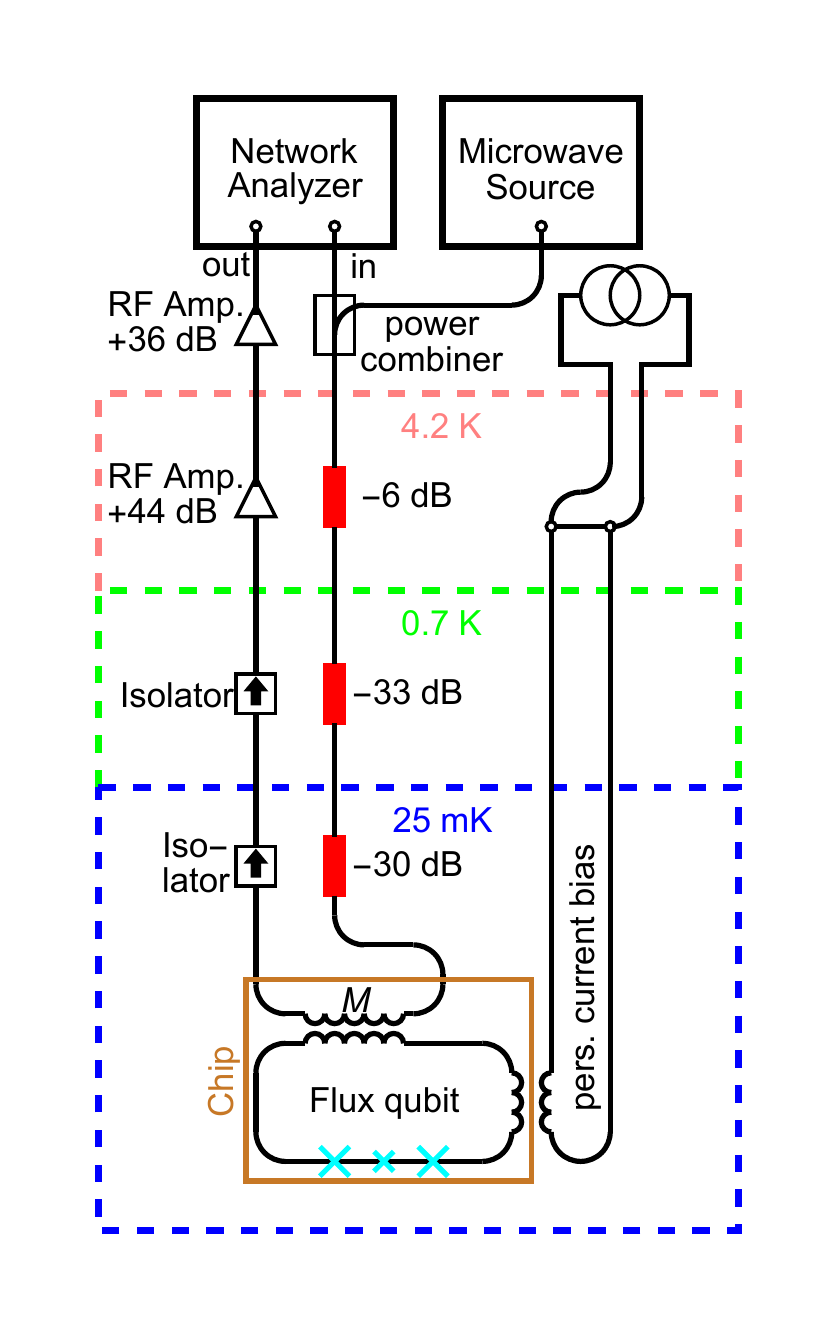}
\caption{Experimental setup for measuring the microwave transmission through the sample chip in a dilution refrigerator with base temperature of $25\,\milli\kelvin$ using a network analyzer. The red boxes display attenuators.}
\label{fig_cryosetup}
\end{figure}

All measurements are conducted in the dilution refrigerator at a base temperature of approximately $25\,\milli\kelvin$. A schematic diagram of the experimental setup with the sample under study and the different circuit elements placed at different temperature levels is shown in Fig.\,\ref{fig_cryosetup}. A large magnetic shielding is obtained by three mu-metal shields at room temperature, a cryoperm shield  around the $4\,\kelvin$ stage and a superconducting lead shield thermally anchored to the still ($0.7\,\kelvin$). 
The flux bias for the qubit is created by the field of an external coil which can be operated in the persistent current mode and hence applies minimal flux noise to the sample. 
We measure the microwave transmission through the sample using a vector network analyzer. The input line is heavily attenuated at various temperature stages to minimize thermal noise at the sample. 
At the output port a cryogenic HEMT amplifier and a room temperature RF amplifier are used to amplify the signal while two isolators protect the sample from the amplifier noise. 
For two-tone experiments, the microwave transmission through the sample is measured by the vector network analyzer while a second microwave tone is supplied to the input line by an additional microwave source. 

\section{Transmittance of an atom embedded into a 1D transmission line with two partial reflectors}

In this section we present a simple model that can be used to analyze the transmittance through a one-dimensional transmission line containing an atom which is surrounded by two partial reflectors. 
The latter originate from small impedance mismatches between the on-chip superconducting coplanar waveguide transmission line and the off-chip wiring. 
Our model is based on the language of transfer matrices and fully reminiscent of scattering theory in one-dimensional systems. 
We first address the issue of a pseudo-resonator that is created by the two impedance mismatches and compute the corresponding bound states.
The resonant structure related to this resonator is referred to as weak resonance in the main text.
In the next step we couple the resonator to a qubit which is placed in the center of this resonator and acts as a scatterer. 
We then calculate the new transmission and reflection properties of the hybridized system. 
In particular, we show that the qubit is affected by the change in the density of states (DOS) induced by the weak resonator. 
Compared to the ideal transmission line (open space scenario) the effect of the weak resonator can be accounted for by a radiative correction.

\subsection{Transmission and Reflection without Qubit}
We consider a superconducting transmission line with impedance $Z_\text{0}\,{=}\,50\,\ohm$ containing two partial reflectors induced by small impedance mismatches $\Delta\/Z_\text{r}\,{=}\,Z_\text{r}-Z_\text{0}$ at the transition to the off-chip microwave wiring. 
These partial reflectors result in a finite scattering of the propagating microwaves  and can be modeled by scattering or transfer matrices. 
We note that a scattering matrix usually connects the incoming modes $a$ and $a^\prime$ on both sides of the scatterer to the outgoing modes $b$ and $b^\prime$, whereas in the transfer matrix formalism the transfer matrix relates the incoming and outgoing modes $a$ and $b$ on one side of the scatterer to the outgoing and incoming modes $a^\prime$ and $b^\prime$ on the other side. 
In the following we will stick to the transfer matrix formalism. 
The transfer matrix captures the dispersive properties of the partial reflectors and can be written explicitly in terms of frequency-dependent reflection and transmission coefficients. 

The general form of a transfer matrix for a partial reflector can be written as~\cite{Shen&Fan1, Shen&Fan2}

\begin{equation} \label{chi}
\mathcal{X}_{Z_\text{r}} = \left (
\begin{array}{cc}
 \dfrac{1}{t^*_{Z_\text{r}}} & - \dfrac{r^*_{Z_\text{r}}}{t^*_{Z_\text{r}}} \\[0.4cm]
- \dfrac{r_{Z_\text{r}}}{t_{Z_\text{r}}} &  \dfrac{1}{t_{Z_\text{r}}}
\end{array}
 \right ) ~ \in ~ {\rm SU} (1,1) ~ ,
 \end{equation}
where $t_{Z_\text{r}}$ and $r_{Z_\text{r}}\,{=}\,\Delta\/Z_\text{r}/(Z_0\,{+}\,Z_\text{r})$ are the complex transmission and reflection coefficient, respectively. If the scattering process is lossless, ${|t_{Z_\text{r}}|}^2 + {|r_{Z_\text{r}}|}^2 = 1$. 
The transfer matrix relates the incoming and outgoing wave amplitudes $a(b^\prime)$ and $b(a^\prime)$ on both sides, 
  \begin{equation}
\left (
\begin{array}{c}
a^\prime \\
b^\prime
\end{array}
\right ) = \mathcal{X}_{Z_\text{r}}
\left ( 
\begin{array}{c}
a \\
b
\end{array}
\right ) ~,
\end{equation}
 where from now on we strictly keep a right moving direction for the transfer matrices. 
 
Note that $\mathcal{X} _{Z_r}\,{\in}\,{\rm SU} (1,1)$ is written in a fully symmetric form~\cite{Pendry} with time reversal symmetry and identical transmission and reflection coefficients for waves incident from either side. 
The unitarity of the corresponding scattering matrix~\cite{Haus} implies conditions, i.e.~$ r_{Z_\text{r}}^* t_{Z_\text{r}}\,{+}\,r_{Z_\text{r}} \, t^*_{Z_\text{r}}\,{=}\,0$, for the orthogonality of the respective eigenmodes. 
In particular, the symmetric reflectivity coefficient  is always $\pi$-shifted from transmittance, i.e.~ $r_{Z_\text{r}}/t_{Z_\text{r}}\,{=}\,\myi \times \mathbb{R}$. Taking into account ${|t_{Z_\text{r}}|}^2 + {|r_{Z_\text{r}}|}^2 = 1$, the reflectivity
\begin{equation}
\frac{r_{Z_\text{r}}}{t_{Z_\text{r}}}\,{=}\, \sqrt{1-  \frac{1}{{|t_{Z_\text{r}}|}^2}}
\end{equation}
can always be specified in terms of transmittance.
 
Our model considers a setup of equal partial reflectors connected by a $50\,\ohm$ transmission line forming a pseudo-resonator. 
The transfer matrix $C$ describing this pseudo-resonator is given by
\begin{equation} \label{defineC}
\mathcal{R}\,{=}\,  \mathcal{X}_{Z_\text{r}} \cdot \tau_{\omega_\text{r}} \cdot  \mathcal{X}_{Z_\text{r}} ~,
\end{equation}
where we use
 \begin{equation}
\tau _{\omega_\text{r}} = \left (
\begin{array}{cc}
\mye^{\myi k s} & 0 \\[0.2cm]
0 &  \mye^{-\myi  ks}
\end{array}
 \right )
 \end{equation}
in order to account for the free and lossless propagation of plane waves in between the two partial reflectors separated by a distance $s$. 
 
The distance $s$ between the partial reflectors can be measured in units of the resonant mode wavelength $\lambda_\text{r}$ of the weak resonator. For the fundamental mode we have $s\,{=}\,\lambda_\text{r}/2$. Then, with $k\,{=}\,2\pi/\lambda$ and a linear dispersion $\omega\,{=}\,c\,k$ we can express the optical path $k\cdot\/s$ as

 \begin{equation}
 k\cdot\/s\,{=}\,\pi\frac{\lambda_{\rm r}}{\lambda}\,{=}\,\pi\frac{\omega}{\omega_\text{r}}.
 \end{equation}
 
For the transfer matrix $\mathcal{R}$ we then obtain the well known result of the transmission through a double tunneling barrier:
\begin{equation} 
\mathcal{R} = \left (
\begin{array}{cc}
 \dfrac{1}{{T_{\rm r}}^*} & - \dfrac{{R_{\rm r}}^*}{{T_{\rm r}}^*} \\[0.4cm]
- \dfrac{R_{\rm r}}{T_{\rm r}} &  \dfrac{1}{T_{\rm r}}
\end{array}
 \right ) ~ \in ~ {\rm SU} (1,1) ~ ,
 \end{equation}
 where
\begin{equation} \label{TC}
T_{\rm r}\,{=}\,\frac{\mye ^{\myi ks} ~t_{Z_\text{r}} ^2}{1\,{-}\, \mye ^{2 \myi ks} ~ r_{Z_\text{r}} ^2}
~~~~~{\rm and}~~~~~
R_{\rm r} = \frac{t_{Z_\text{r}}}{t^*_{Z_\text{r}}} ~\frac{\mye ^{2 \myi ks} ~r_{Z_\text{r}}\,{-}\,r^*_{Z_\text{r}}}{1\,{-}\, \mye ^{2 \myi ks} \, r_{Z_\text{r}} ^2} ~.
\end{equation}
We immediately see that reflection and transmission are $\pi$-phase-shifted, i.e.,
\begin{equation} \label{RC/TC}
\frac{R_{\rm r}}{T_{\rm r}}\,{=}\,2 \myi ~\frac{ {\rm Im} \left (\mye ^{\myi ks} \, r_{Z_\text{r}} \right )}{{| t_{Z_\text{r}} |}^2} ~.
\end{equation}
This implies ${R_{\rm r}}^* T_{\rm r}\,{+}\,R_{\rm r} \, {T_{\rm r}}^*\,{=}\, 0$ and allows us to introduce normalizable bound states for the pseudo-resonator. Considering waves incident from opposite directions, we seek two scattered waves that are orthogonal, i.e.~\cite{Haus}
\begin{equation}
\int _{-\infty}^{\infty} {\rm d}x ~{\phi _{\rm r}}^* (x) \, \phi _{\rm l} (x)\,{=}\,0 ~.
\end{equation}

If we place the origin $x\,{=}\,0$ at the center of the resonator, we find for the left incoming wave with uniform scaling $L$:
\begin{equation} \label{phil}
	\phi _{\rm l} (x)= \frac{1}{\sqrt{2 L}} \left \{ \begin{aligned}
	\mye ^{\myi k ( x + s / 2)} + R_{\rm r} ~\mye ^{-\myi k ( x + s / 2)} &,~& x < - \frac{s}{2} \\[0.2cm]
	 \frac{T_{\rm r}}{t_{Z_\text{r}}} ~\mye ^{\myi k ( x - s / 2)} + r_{Z_\text{r}} \frac{T_{\rm r}}{t_{Z_\text{r}}} ~\mye ^{-\myi k ( x - s / 2)} &,~& - \frac{s}{2} < x < \frac{s}{2} \\[0.2cm]
	T_{\rm r} ~ \mye ^{\myi k ( x - s / 2)} &,~& x > \frac{s}{2} \\
	\end{aligned}
	\right . ~.
\end{equation}
and similar for the right incoming wave:
\begin{equation} \label{phir}
	\phi _{\rm r} (x) = \frac{1}{\sqrt{2 L}} \left \{ \begin{aligned}
	T_{\rm r} ~ \mye ^{-\myi k ( x + s / 2)} &,~& x < -\frac{s}{2} \\[0.2cm]
	r_{Z_\text{r}} \frac{T_{\rm r}}{t_{Z_\text{r}}} ~\mye ^{\myi k ( x + s / 2)} + \frac{T_{\rm r}}{t_{Z_\text{r}}} ~\mye ^{-\myi k ( x + s / 2)} &,~& - \frac{s}{2} < x < \frac{s}{2} \\[0.2cm]
	 R_{\rm r} ~\mye ^{\myi k ( x - s / 2)} + \mye ^{-\myi k ( x - s / 2)}&,~& x > \frac{s}{2} \\
	\end{aligned}
	\right . ~ .
\end{equation}

The latter can be formulated in terms of the symmetric and anti-symmetric superpositions, i.e.
\begin{equation}
\phi_\pm (x) = \frac{1}{\sqrt{2}} \left [ \phi _{\rm l} (x)  \pm \phi _{\rm r} (x)\right ] ~.
\end{equation}

We are only interested in the intra-resonator solutions $\psi_\pm (x)\,{\equiv}\,\phi_\pm (x)$ with $-s/2 < x < s/2$. From Eqs.~(\ref{phil}, \ref{phir}) we hence obtain
\begin{equation}
	\psi_- (x) =  \myi  \frac{1}{\sqrt{L}} \frac{T_{\rm r}}{t_{Z_\text{r}}} \left ( \mye ^{-\myi k s / 2} - \mye ^{\myi k s / 2} \, r_{Z_\text{r}} \right ) \sin (k x) ~~~ , ~~~
	\psi_+ (x) =  \frac{1}{\sqrt{L}} \frac{T_{\rm r}}{t_{Z_\text{r}}} \left ( \mye ^{-\myi k s / 2} + \mye ^{\myi k s / 2} \, r_{Z_\text{r}} \right ) \cos (k x) ~,
\end{equation}
which yields
\begin{equation} \label{LDOS}
 {\left | \psi_- (x) \right |} ^2 =  \frac{{| t_{Z_\text{r}} |} ^2}{{\left | 1 + \mye ^{\myi k s} \,r_{Z_\text{r}} \right |}^2} \frac{\sin ^2 (x) }{L} ~~~{\rm and}~~~
{ \left | \psi_+ (x) \right |} ^2 =  \frac{{| t_{Z_\text{r}} |} ^2}{{\left | 1 - \mye ^{\myi k s} \,r_{Z_\text{r}} \right |}^2} \frac{\cos ^2 (x) }{L} ~.
\end{equation}

\subsection{Transmission and Reflection with Qubit}
The most general form for a transfer matrix with time reversal symmetry and symmetric reflectance $r_{\rm q}$ and transmittance $t_{\rm q}$ can be written as \cite{Pendry}
 \begin{equation}
Q = \left (
\begin{array}{cc}
t_{\rm q} - \dfrac{{r_{\rm q}}^2}{t_{\rm q}} & \dfrac{r_{\rm q}}{t_{\rm q}} \\[0.4cm]
-\dfrac{r_{\rm q}}{t_{\rm q}} &  \dfrac{1}{t_{\rm q}}
\end{array}
 \right ) ~ .
 \end{equation}
 
We note that the general transfer matrix $Q$ may have lower symmetry than the transfer matrix in Eq.~(\ref{chi}) if it does not conserve probabilities, i.e.,${|t_{\rm q}|}^2 + {|r_{\rm q}|}^2 \not = 1$. 
This allows us to consider possible dephasing effects. 
This is particularly useful when the transfer matrix $Q$ is applied to to model the scattering event with a qubit.

In order to obtain the transfer matrix $\mathcal{T}$ for coupled qubit-resonator-system, we extend Eq.~(\ref{defineC}) and place the qubit in the centre of the resonator. The total transfer matrix then reads as 
\begin{equation}
\mathcal{T}\,{=}\,  \mathcal{X} _{Z_r} \cdot \tau_{2\omega_{r}} \cdot Q \cdot \tau_{2\omega_{r}} \cdot \mathcal{X} _{Z_r} ~.
\end{equation}
Without specifying transmittance $t_{\rm q}$ and reflectance $r_{\rm q}$ for the qubit, we obtain
 \begin{equation}
\mathcal{T} = \left (
\begin{array}{cc}
T_{\rm t} - \dfrac{{R_{\rm t}}^2}{T_{\rm t}} & \dfrac{R_{\rm t}}{T_{\rm t}} \\[0.4cm]
-\dfrac{R_{\rm t}}{T_{\rm t}} &  \dfrac{1}{T_{\rm t}}
\end{array}
 \right )
 \end{equation}
 with
\begin{equation} \label{TH}
 T_{\rm t} = t_{\rm q} ~\frac{\mye ^{\myi k s} ~t^2_{Z_\text{r}}}{1 - 2 r_{\rm q} r_{Z_\text{r}} ~ \mye ^{\myi k s} -  r ^2 _{Z_\text{r}} \left ( {t_{\rm q}} ^2- {r_{\rm q}}^2 \right ) ~\mye ^{2 \myi k s} }
 \end{equation}
 and
 \begin{equation}
 R_{\rm t} = - \frac{t_{Z _\text{r}}}{t^* _{Z_\text{r}}}  ~ \frac{r^* _{Z_\text{r}} - r_{\rm q} \left (1 + {| r _{Z_\text{r}}|} ^2 \right ) ~\mye ^{\myi k s} - r _{Z_\text{r}} \, \left ( {t_{\rm q}} ^2- {r_{\rm q}}^2 \right ) ~\mye ^{2 \myi k s}  }{1 - 2 r_{\rm q} r _{Z_\text{r}} ~ \mye ^{\myi k s} -  r ^2 _{Z_\text{r}} \left ( {t_{\rm q}} ^2- {r_{\rm q}}^2 \right ) ~\mye ^{2 \myi k s} } ~.
  \end{equation}

The remaining task is to derive expression for the transmission and reflection coefficient of the qubit. We start from Ref.~\onlinecite{astafiev_fluorescent} and assume low power and a unit coupling efficiency between qubit and the line field. Then, for a qubit that emits with the same phase in both directions ($t_{\rm q} - r_{\rm q} =1$), we have for any arbitrary relaxation rate $\Gamma_\text{1}$ and dephasing rate $\Gamma_\varphi$
\begin{equation} \label{t}
{t_{\rm q} }^{-1} = 1 + \frac{1}{2} \frac{\Gamma_\text{1}}{\Gamma_\varphi - \myi \Delta}
\end{equation}
with $\Delta = \omega - \omega_{\rm q}$ being the detuning from the qubit transition frequency $\omega_{\rm q}$. 

Note that the transmission coefficient given by Eq.~(\ref{t}) is applicable to describe the resonance fluorescence of the qubit in open space. 
In this case we can assign a physical meaning to $\Gamma_\text{1}$ and $\Gamma_\varphi$: $\Gamma_\text{1}$ is the spontaneous relaxation rate and $\Gamma_\varphi$ the pure dephasing rate \cite{myfootnote}.

In order to show that this meaning can be restored also for a transmission line containing two partial reflectors, we calibrate the transmittance in Eq.~(\ref{TH}) for the hybridized system with respect to the transmittance of the qubit-free case in Eq.~(\ref{TC}). 
For the calibrated transmittance $T_{\rm cal}$, we obtain
\begin{equation}
{T_{\rm cal} }^{-1}\,{=}\,1\,{+}\,\frac{1}{2} \frac{\Gamma_\text{1} \, z}{\Gamma_\varphi - \myi \Delta}~\Rightarrow~T_{\rm cal}\,{=}\,\frac{2\,(\Gamma_\varphi\,{-}\,i\,(\omega\,{-}\,\omega_\text{q}) )}   {\Gamma_\text{1}\,z\,{+}\,2\,\Gamma_\varphi\,{-}\,2\,i\,(\omega\,{-}\,\omega_\text{q})}
\end{equation}
with 
\begin{equation}
z\,=\,\frac{1\,{+}\, \mye ^{\myi ks } \, r_{Z_\text{r}} }{1\,{-}\,\mye ^{ \myi ks } \,  r_{Z_\text{r}}}~.
\end{equation}
Evidently, the transmission coefficient of the coupled qubit-resonator-system is obtained from the one for a qubit in open space by multiplying with a radiative correction factor.

With Eq.~(\ref{RC/TC}) and Eq.~(\ref{LDOS}) we can rewrite the latter in terms of its real and imaginary part, i.e.
\begin{equation}
z\,{=}\,\left (1\,{+}\, \frac{R_{\rm r}}{T_{\rm r}} \right ) ~\frac{{\left | \psi_+ (0) \right |} ^2}{~{ | \tilde{\psi}_+ (0)  |} ^2} ~.
\end{equation}

\begin{figure}[h!]
\includegraphics{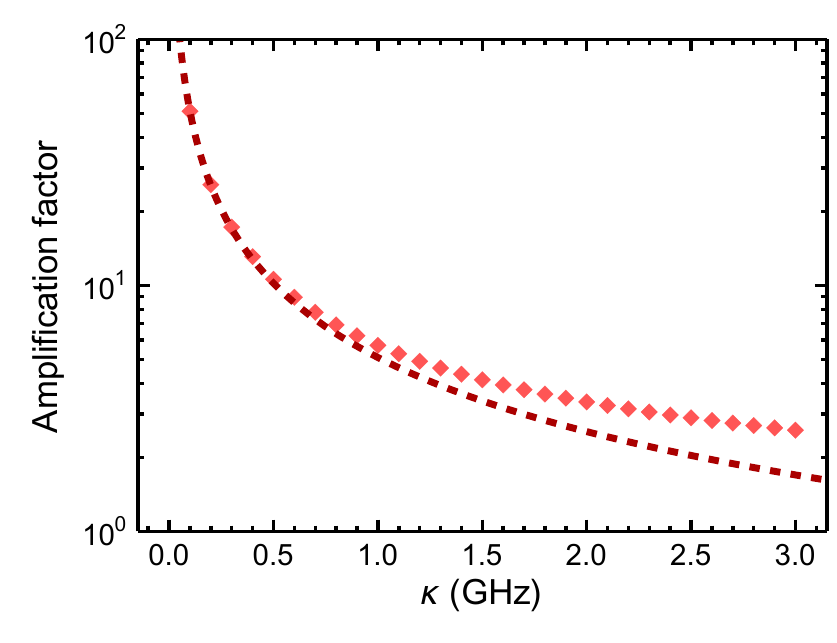}
\caption{Amplification factor of the spontaneous emission rate due to Purcell effect (dashed line) or our model (diamonds).}
\label{fig_purcell}
\end{figure}

We stress that the qubit couples only to the symmetric cavity mode $\psi_+ (x)$ since the anti-symmetric mode vanishes completely at the position of the qubit,  i.e.~$\psi_-(0)\,{=}\,0$. The real part of $z$ is consistent with the calibrated DOS at the position of the qubit inside the cavity,  i.e.~$\psi_+ (x)\,{=}\,\tilde{\psi}_+ (x)$  with $t\,{\rightarrow}\,1$.  With $R_{\rm c}/ T_{\rm c}\,{=}\, \myi \times \mathbb{R}$, the imaginary part of $z$ marks a {\sl Lamb}-shift for the resonance of the calibrated transmittance $T_{\rm cal}$.

In the remainder of this section, we compare our model to the Purcell effect where the density of states of the bath is assumed to be resonator-like. The Purcell factor measures the change in the spontaneous emission due to the altered density of states. The one-dimensional representation of this factor reads\cite{gambetta_purcellprotection}

\begin{equation}
\label{purcellfac}
z_\text{Purcell}\,{=}\,\frac{1}{\pi}\frac{\kappa\,\omega}{\Delta^2\,{+}\,(\kappa/2)^2}~,
\end{equation}
where $\kappa$ is the decay rate of the resonator. In Fig.~\ref{fig_purcell}, we compare our model to Eq.~(\ref{purcellfac}). As expected, for small $\kappa$ values our model tends towards the Purcell factor.

\section{Spin-boson model of a qubit in an open transmission line}

We start with the interaction Hamiltonian of a flux qubit coupled to a transmission line~\cite{peropadre_dynuscoupling}
\begin{equation}
\label{eq_Hint_1}
H_\text{int}\,=\,\hbar\,\sigma_x\,\sum_k\,\left(g_{\text{k}} a_\text{k}^\dagger+g_\text{k}^*a_\text{k}\right).
\end{equation}
Here, $\sigma_{x}$ is a Pauli spin operator and $g_\text{k}$ is the coupling to the field mode $(a_\text{k} + a_\text{k}^\dagger)$. 
On the other hand, for an inductive coupling we can write 
\begin{equation}
H_\text{int}\,=\,M\,I_\text{line}\,I_\text{p}~,
\end{equation}
where $M$ is the mutual inductance and $I_\text{line}$ is the vacuum current of the line.
The vacuum current reads as
\begin{equation}
\label{eq_Hint_2}
I_\text{line}\,=\,i\,\sum_k\,\frac{1}{Z}\,\sqrt{\frac{\hbar}{2\,l_\text{0}}}\,\sqrt{\frac{\omega_\text{k}}{L}}\,(a_\text{k}\,e^{-i\,k\,x} + a_\text{k}^\dagger\,e^{i\,k\,x})~.
\end{equation}
Here, $l_0$ is the self-inductance per unit length, $Z$ is the line impedance and $L$ is the mode volume.
Using Eq.\,(\ref{eq_Hint_1}-\ref{eq_Hint_2}), we obtain
\begin{equation}
\label{eq_gk2}
g_\text{k}\,=\,G\,\sqrt{\frac{\omega_\text{k}}{L}}\,e^{-i\,k\,x}
\end{equation}
with 
\begin{equation}
\label{capG}
G\,=\,i\,M\,I_\text{p}\,\frac{1}{Z}\,\sqrt{\frac{\hbar}{2\,l_\text{0}}}~. 
\end{equation}

In the next step, we want to model a situation where the qubit only couples to one specific mode $\omega_\text{k}$ equivalent to resonator coupling. In closed boundary conditions
\begin{equation}
\label{eq_gk}
g_\text{k}\,=\,G\,\sqrt{\frac{\omega_\text{k}}{L}}\,\sin(k x)
\end{equation}
with the normalization $L\,=\,\lambda/4$.

In the spin-boson model, the spectral function $J(\omega)$ describing a quantum two-level system coupled to the environment consisting of an infinite number of bosonic degrees of freedom is defined as
\begin{equation}
J(\omega)\,=\,2\,\pi\,\sum_k g_\text{k} g_\text{k}^*\,\delta(\omega-\omega_\text{k})~.
\end{equation}
Using Eq.\,(\ref{eq_gk}) and moving to a continuum description, we find
\begin{equation}
\label{JofG}
J(\omega)\,{=}\,\frac{G G^*}{v} D(\omega)\,\omega~,
\end{equation}
where $v$ is the phase velocity of the wave and $D(\omega)$ is the density of states in the line.
In order to compare the transmission line coupling with the resonator coupling, we assume that the transmission line couples only to a single mode. Using $D(\omega)\,=\,2$, we find
\begin{equation}
|g_\text{tra}|\,=\,\sqrt{\frac{\Gamma_\text{1}\,\omega}{\pi}}~.
\end{equation}

Sometimes it is convenient to express the Ohmic spectral function in terms of the dimensionless Kondo parameter~\cite{peropadre_dynuscoupling} $J(\omega)\,{=}\,2\,\pi\,\alpha\,\omega$. From Eq.~(\ref{JofG}) and Eq.~(\ref{capG}) we can derive
\begin{equation}
\alpha\,{=}\,\frac{G G^*}{2\,\pi\,v} D(\omega)\,{\propto}\,M^2~.
\end{equation}

\section{Acknowledgements}

This work is supported by the German Research
Foundation through SFB 631 and FE 1564/1-1, EU projects CCQED,
PROMISCE and SCALEQIT, the doctorate program ExQM of the Elite Network of Bavaria,
Spanish MINECO Project FIS2012-33022 and CAM Research Network QUITEMAD+.
E. S. acknowledges support from Basque Government IT472-10; Spanish MINECO FIS2012-36673- C03-02; UPV/EHU UFI 11/55; PROMISCE and SCALEQIT EU projects. E.S. acknowledges the hospitality of Walther-Mei{\ss}ner-Institut and Institute for Advanced Study at TUM.

\bibliography{bib}